\documentstyle[galley,graphicx,epsf]{mn}
\newif\ifAMStwofonts
\ifoldfss
  \ifCUPmtlplainloaded \else
    \NewTextAlphabet{textbfit} {cmbxti10} {}
    \NewTextAlphabet{textbfss} {cmssbx10} {}
    \NewMathAlphabet{mathbfit} {cmbxti10} {} % for math mode
    \NewMathAlphabet{mathbfss} {cmssbx10} {} %  "   "    "
  \fi
  \ifAMStwofonts
    \ifCUPmtlplainloaded \else
      \NewSymbolFont{upmath} {eurm10}
      \NewSymbolFont{AMSa} {msam10}
      \NewMathSymbol{\upi}     {0}{upmath}{19}
      \NewMathSymbol{\umu}     {0}{upmath}{16}
      \NewMathSymbol{\upartial}{0}{upmath}{40}
      \NewMathSymbol{\leqslant}{3}{AMSa}{36}
      \NewMathSymbol{\geqslant}{3}{AMSa}{3E}

      \let\leq=\leqslant \let\le=\leqslant
       
    \fi
  \fi
\fi % End of OFSS

\ifnfssone
  \newmathalphabet{\mathit}
  \addtoversion{normal}{\mathit}{cmr}{m}{it}
  \addtoversion{bold}{\mathit}{cmr}{bx}{it}
  \newmathalphabet{\mathbfit} % math mode version of \textbfit{..}
  \addtoversion{normal}{\mathbfit}{cmr}{bx}{it}
  \addtoversion{bold}{\mathbfit}{cmr}{bx}{it}
  \newmathalphabet{\mathbfss} % math mode version of \textbfss{..}
  \addtoversion{normal}{\mathbfss}{cmss}{bx}{n}
  \addtoversion{bold}{\mathbfss}{cmss}{bx}{n}
  \ifAMStwofonts
    \ifCUPmtlplainloaded \else
      \UseAMStwoboldmath
      \makeatletter
      \new@mathgroup\upmath@group
      \define@mathgroup\mv@normal\upmath@group{eur}{m}{n}
      \define@mathgroup\mv@bold\upmath@group{eur}{b}{n}
      \edef\UPM{\hexnumber\upmath@group}
      \new@mathgroup\amsa@group
      \define@mathgroup\mv@normal\amsa@group{msa}{m}{n}
      \define@mathgroup\mv@bold\amsa@group{msa}{m}{n}
      \edef\AMSa{\hexnumber\amsa@group}
      \makeatother
      \mathchardef\upi="0\UPM19
      \mathchardef\umu="0\UPM16
      \mathchardef\upartial="0\UPM40
      \mathchardef\leqslant="3\AMSa36
      \mathchardef\geqslant="3\AMSa3E

      \let\leq=\leqslant \let\le=\leqslant

    \fi
  \fi
\fi % End of NFSS release 1

\ifnfsstwo
  \DeclareMathAlphabet{\mathbfit}{OT1}{cmr}{bx}{it}
  \SetMathAlphabet\mathbfit{bold}{OT1}{cmr}{bx}{it}
  \DeclareMathAlphabet{\mathbfss}{OT1}{cmss}{bx}{n}
  \SetMathAlphabet\mathbfss{bold}{OT1}{cmss}{bx}{n}
  \ifAMStwofonts
    \ifCUPmtlplainloaded \else
      \DeclareSymbolFont{UPM}{U}{eur}{m}{n}
      \SetSymbolFont{UPM}{bold}{U}{eur}{b}{n}
      \DeclareSymbolFont{AMSa}{U}{msa}{m}{n}
      \DeclareMathSymbol{\upi}{0}{UPM}{"19}
      \DeclareMathSymbol{\umu}{0}{UPM}{"16}
      \DeclareMathSymbol{\upartial}{0}{UPM}{"40}
      \DeclareMathSymbol{\leqslant}{3}{AMSa}{"36}
      \DeclareMathSymbol{\geqslant}{3}{AMSa}{"3E}

      \let\leq=\leqslant \let\le=\leqslant

    \fi
  \fi
\fi % End of NFSS release 2

\ifCUPmtlplainloaded \else
  \ifAMStwofonts \else % If no AMS fonts
    \def\upi{\pi}
    \def\umu{\mu}
    \def\upartial{\partial}
  \fi
\fi

\begin{document}
\title{The R Coronae Borealis stars - carbon abundances from forbidden carbon lines}
\author[G. Pandey et al. ]
       {Gajendra Pandey$^1$, David L. Lambert$^1$, N. Kameswara Rao$^2$,
Bengt Gustafsson$^3$, 
\newauthor
Nils Ryde$^3$, David Yong$^1$\\
       $^1$Department of Astronomy, University of Texas, Austin, TX 78712-1083, USA\\
       $^2$Indian Institute of Astrophysics, Bangalore 560034, India\\
       $^3$Astronomiska observatoriet, Uppsala, Sweden S-751-20\\}
\date{Accepted .
      Received ;
      in original form 2002 }

\pagerange{\pageref{firstpage}--\pageref{lastpage}}
\pubyear{2001}

\maketitle

\label{firstpage}

\begin{abstract}

Spectra of several R Coronae Borealis (RCB) stars at 
maximum light were examined for the
[C\,{\sc i}] 9850\AA\ and 8727\AA\ absorption lines. The 9850\AA\
line is variously blended with a Fe\,{\sc ii} and CN lines but
positive identifications of the [C\,{\sc i}] line are
made for R\,CrB and SU\,Tau. The 8727\AA\ line is detected in
the spectrum of the five stars observed in this wavelength region.
Carbon abundances are derived from the
[C\,{\sc i}] lines using the model atmospheres and atmospheric
parameters used by Asplund et al. (2000).

%Since photoionization of neutral carbon atoms from high
%excitation levels is the dominant source
%of continuous opacity, the predicted strength of  weak
%high excitation  C\,{\sc i}  
%lines are independent of the input C abundance, and insensitive to the
%adopted atmospheric parameters. 
Although the observed
strength of a  C\,{\sc i} line is constant from cool to hot
RCB stars, the strength is weaker than predicted by
an amount equivalent to a factor of four reduction of a line's $gf$-value.
Asplund et al. dubbed this `the carbon problem' and discussed 
possible solutions. 

The [C\,{\sc i}] 9850\AA\
line seen clearly in R\,CrB and SU\,Tau confirms the magnitude of
the carbon problem revealed by the C\,{\sc i} lines. The
[C\,{\sc i}] 8727 \AA\ line measured in five stars shows an enhanced carbon
problem. The $gf$-value required to fit the observed [C\,{\sc i}] 8727 \AA\ line is a
factor of 15 less than the well-determined theoretical $gf$-value.
We suggest that the carbon problem
for all lines  may be alleviated to some extent by a
chromospheric-like temperature rise in these stars. 
The rise far exceeds that predicted by our non-LTE
calculations, and requires a substantial deposition of mechanical
energy.

\end{abstract}


\section{Introduction}

The rare class of R Coronae Borealis (RCB) variables continues to present
fascinating puzzles. Two defining characteristics of these supergiants
are the  unpredictable fading of the star and the presence of helium, not hydrogen,
as the most common constituent of the stellar atmosphere.
The fadings are generally assumed to occur when a cloud of carbon
soot forms on the line of sight to the star, but the formation
process, and initial location and evolution of the cloud encompass
unanswered questions. How a normal star or a pair of stars evolves to
possess a H-deficient atmosphere has challenged theorists and
observers alike. A leading idea is that an RCB star is the result of
a merger of a He white dwarf with  a C-O white dwarf. To develop observational
tests of this and other ideas, one requires accurate estimates of the
elemental abundances for a representative sample of the RCB stars.

In attempting to supply this data, Asplund et al. (2000) uncovered yet
another interesting puzzle which is discussed below. 
In RCB stars, as in normal stars, the strength of an absorption line is
influenced by two principal factors: the ratio of the line to continuous
opacity and the temperature gradient through the region of the atmosphere
in which the line is formed.
It has been suggested that photoionization of neutral carbon from high-lying levels
appears to be the dominant source of continuous absorption across the visible 
spectrum (Searle 1961).
A characteristic of RCB star spectra is a rich collection of
absorption lines of neutral carbon from levels only
slightly less excited than those contributing to the
continuous opacity.
Since the line and continuous opacity originate from similar excited
levels of the neutral carbon atoms, their opacity ratio is 
quite insensitive to fundamental parameters of the stars such as
effective temperature, surface gravity, and the fractional carbon
abundance (effectively, the C/He ratio once it exceeds 1\% $-$ see also Sch\"{o}nberner 1975; 
Cottrell \& Lambert 1982). For weak lines, this
insensitivity extends to differences in the turbulent 
velocities.  
%(Sch\"{o}nberner 1975; Cottrell \& Lambert 1982).
%In short, since photoionization of neutral carbon atoms from high
%excitation levels is the dominant source
%of continuous opacity, the predicted strength of weak
%high excitation  C\,{\sc i}
%lines are independent of the input C abundance, and insensitive to the
%adopted atmospheric parameters. 

This is observed in the RCB stars.
A C\,{\sc i} line retains its equivalent width even as `metal' lines 
may vary considerably from star to star (see Figure 1 in Rao \& Lambert 1996). 
However, the
predicted equivalent widths of the C\,{\sc i} lines exceed the observed
equivalent widths by a considerable factor.
Expressed as the change in a line's
$gf$-value needed to bring the predicted in tune with the observed
equivalent width, the disagreement amounts to a factor of four. Asplund et al.
dubbed this `the carbon problem'.

The principal purpose of this paper is to add a
new clue to the solution to the carbon problem. That dimension
is offered by the [C\,{\sc i}]  8727 \AA\ and 9850 \AA\
lines and the estimate of the carbon abundance that they provide.
We propose to investigate this aspect in more stars and to see
whether a solution can be obtained for this puzzle. The present paper
provides the results of these observations which, to our surprise, show
an enhanced carbon problem. 
%These lines from the ground parent configuration
%of the neutral carbon atom and the permitted high-excitation C\,{\sc i}
%lines from excited configurations
%used by Asplund et al. differ in several major ways which are discussed.
% Consideration of the
%advantages and disadvantages of forbidden versus permitted lines
%does not indicate a clear balance in favor of the former.
%It was hoped that the combination of the lines would shed new light on the 
%carbon problem. 
%The opportunity is taken also to  reexamine Asplund et al.'s conclusion
% based on measurements of the
% 9850 \AA\  line that it provides a carbon abundance which matches
%the input carbon abundance of a model, i.e.,   the carbon problem
%does not arise for the 9850 \AA\ line.

\section{Observations}
High-resolution optical spectra of eight RCB stars (Table 1) were obtained at the 
W. J. McDonald Observatory's 2.7-m telescope with the coud\'{e} cross-dispersed echelle
spectrograph (Tull et al. 1995). 
%Eight RCBs were observed at a resolving power (R =$\lambda/\Delta\lambda$) 
%of 60,000. Two bright RCBs were observed at a
%resolving power of 120,000. 
In addition, $\gamma$ Cyg, a normal F-type supergiant was also observed (Table 1)
at R = 120,000.
%The detector was a Tektronix 2048 $\times$ 2048 CCD. The R = 60,000 spectra covered the wavelength
%range from 3800\AA\ to 10000\AA, but with incomplete spectral coverage longward of about
%5500\AA. The R = 120,000 spectra cover the desired 8727\AA\ and the 9850\AA\ regions.
%A Th-Ar hollow cathode lamp was observed either just prior to or just after 
%exposures of the programme stars to provide wavelength calibration. The pixel-to-pixel variation
%in the sensitivity of the CCD was removed by the exposures obtained of a halogen lamp. Typical
%exposure times of our programme stars were not more than 30 minutes, and exposures were
%co-added to improve the signal-to-noise ratio of the final spectrum. The
%signal-to-noise (S/N) per pixel in the continuum of the co-added spectra at 8727\AA\ and at 9850\AA\
%are given in Table 1. 
%For each observing run an early-type rapidly rotating star
%was observed to facilitate removal of the telluric absorption lines from
%the spectra of the programme stars. 
%We have used the Image Reduction and
%Analysis Facility (IRAF) software packages to reduce the spectra. 
The telluric absorption lines
from the spectra of the programme stars were removed interactively 
%using the task $telluric$ within IRAF 
using early-type rapidly rotating stars.
We have used the Image Reduction and
Analysis Facility (IRAF) software packages to reduce the spectra, and the task $telluric$ within
IRAF to remove the telluric absorption lines.

The Th-Ar hollow cathode lamp fails to provide lines in the 9850\AA\ region for wavelength calibration.
Plenty of Th-Ar lines are available shortward of 9670\AA\ to provide a wavelength solution
to an accuracy of one tenth of a pixel. We apply this solution for the wavelengths longer
to 9670\AA. To check the accuracy of the wavelength calibration in the 9850\AA\ region, the measured
wavelengths of the atmospheric H$_{2}$O lines are compared with those measured from the
solar spectrum (Swensson, Benedict, Delbouille \& Roland 1970).
%%; we have used the mean wavelength,
%%given in column 3 by these authors, obtained from measures of at least two tracings of the same spectral
%%interval. 
We measured the wavelengths of 11 atmospheric H$_{2}$O lines finding agreement to 
0.003\AA\ $\pm$ 0.02\AA\ with those measured from the solar spectrum.
%To check the wavelength calibration of the R = 120,000 and, R = 60,000 spectra,
%the radial velocites of the stars obtained from the absorption lines shorter of 9670\AA\
%were compared with those obtained from the absorption lines in the 9850\AA\ region. Agreement
%is found within the expected uncertainties.

Sample portions of R = 120,000 and, R = 60,000 spectra are shown in Figures 1, and 2. 
Note that, the spectra of the RCB stars are
fully resolved at R = 60,000. All spectra are aligned to the rest wavelengths
of N\,{\sc i} lines which fall in the wavelength regions.
Stellar lines were identified using the Revised Multiplet Table (Moore 1972), 
tables of spectra of H, C, N, and O (Moore 1993), Kurucz's list\footnote{http://kurucz.harvard.edu},
the Vienna Atomic Line Database\footnote{http://www.astro.univie.ac.at/$\sim$vald},
the new Fe\,{\sc i} multiplet table (Nave et al. 1994), and a spectrum of $\gamma$ Cyg, 
a normal F-type supergiant, as a reference.

All stars except 
V482\,Cyg were observed at maximum light. Our spectrum of V482\,Cyg was
taken when the star was in a shallow minimum in July 1996,
about 2 magnitudes below maximum light. It took the star about one year to return
to maximum light.

\begin{table*}
\centering
\begin{minipage}{90mm}
\caption{Table of observations; the stars are ordered by their effective temperatures decreasing from top to bottom}
\begin{tabular}{lrrrrr}
\hline
\multicolumn{1}{c}{Star}&\multicolumn{1}{c}{Date}&\multicolumn{1}{c}{R = }&\multicolumn{1}{c}{S/N}&
\multicolumn{1}{c}{S/N} &\multicolumn{1}{c}{$V_{rad}$}\\
&&\multicolumn{1}{c}{$\lambda/\Delta\lambda$}& \multicolumn{1}{c}{at 8727 \AA} 
& \multicolumn{1}{c}{at 9850 \AA} & \multicolumn{1}{c}{km s$^{-1}$}\\
\hline
RCB stars:  &                &          &            &     & \\
XX Cam     &   9 Oct 1997   &  60,000  &  ...       & 185 &11\\
XX Cam     &  26 Jan 1998   &  60,000  & ...   & 120 &13\\
XX Cam     &  17 Nov 2002  &  60,000  & ... & 186&11\\
XX Cam     &  17 Nov 2002  &  60,000  & 278 & ...&...\\
RY Sgr     &  22 June 1997  &  60,000  & ...  & 170 & $-$30\\
RY Sgr     &  31 July 2002  &  120,000  & 300  & 200&$-$12\\
UV Cas     &  24 July 1996  &  60,000  & ...   & 140 &$-$32\\
UV Cas     &  15 Nov 2002  &  60,000  &  ...  & 120 &$-$32\\
UV Cas     &  15 Nov 2002  &  60,000  & 178   & ...&...\\
VZ Sgr     &  18 Aug 1999   &  60,000  & ...   & 80 &245\\
R CrB      &  17 June 1995  &  60,000  &   340                & 180&20\\
R CrB      &   7 Aug 1995  &  60,000  &    ...               & 110&22\\
R CrB      &  20 June 1997  &  60,000  &   ...      & 185 &21\\
R CrB      &  31 July 2002  &  120,000  & 400  & 200 &19\\
SU Tau     &  15 Nov 2002   &  60,000  & ...   & 100    &43\\
SU Tau     &  15 Nov 2002   &  60,000  & 180  & ...  &...\\
V482 Cyg   &  24 July 1996  &  60,000  & ...   & 90  &$-$40\\
GU Sgr     &  22 June 1997  &  60,000  & ...   & 95  &$-$45\\
Standard star: &            &          &       &     &   \\
$\gamma$ Cyg& 31 July 2002  &  120,000  &    520            & 230 &...\\
\hline
\end{tabular}
%$^a$The stars are ordered based on their increasing effective temperatures\\
%$^a$covers 8727\AA\ region\\
%$^b$covers the 8727\AA\ and 9850\AA\ regions
\end{minipage}
\end{table*}

\section{Line data and identifications}

\subsection{The [C\,{\sc i}] lines}

%An accurate wavelength and $gf$-value are required
% for each line included in spectrum synthesis. 
The standard reference for the C\,{\sc i} spectrum (Moore 1993)
gives a  predicted wavelength of
8727.126 \AA\ for the [C\,{\sc i}] $2p^2$ $^1$D$_2$ -- $2p^2$ $^1$S$_0$
transition and 
 9850.264 \AA\ for the [C\,{\sc i}]
$2p^2$ $^3$P$_2$ -- $2p^2$ $^1$D$_2$  transition. 
%as computed from energy levels derived from combinations of
%ultraviolet lines between
% levels of the ground configuration and a common upper excited
%level.
% A plausible uncertainty in the predicted position of the line is $\pm$0.02 cm$^{-1}$
%or $\pm$0.019\AA.
 Wavelengths of the forbidden carbon lines have not been measured directly
from laboratory sources.

The predicted wavelength of the 8727 \AA\ line
 is confirmed by the wavelength of the absorption line in the solar
spectrum (Allende Prieto, Lambert, \& Asplund 2002) and the emission line in planetary
nebulae (Liu et al. 1995).
For the 8727 \AA\ line,
we adopt $\log gf$ = $-$8.14 (Galavis, Mendoza \& Zeippen 1997).
 The lower excitation potential is 1.264 eV.
 There is a blending Fe\,{\sc i} line for which we adopt the
$\log gf$ = $-$4.4 (Allende Prieto, Lambert, \& Asplund 2002), but
this line is not a significant
contributor to the spectrum of an RCB star.

For the 9850 \AA\ line, we adopt the predicted wavelength.
 There is supporting evidence
from our spectra for this wavelength.
 The 2002 July spectra of
RY\,Sgr show the 9850 \AA\ and 8727 \AA\ [C\,{\sc i}] lines in emission
(see Figures 1 and 2). Adopting 8727.126 \AA\ as the rest
wavelength for the latter emission and assuming the two lines
have the same velocity, we find the astrophysical wavelength
of  the 9850 \AA\ emission is close to the predicted wavelength.
Liu et al. (1995) report, however,
 a rest wavelength of 9850.36 $\pm$ 0.01\AA\ from detections
of the line in four planetary nebulae. 
The 0.1\AA\ or 3km s$^{-1}$
 difference between the wavelength predicted from
energy levels and that observed from planetary nebulae does not
seriously impact derivation of the carbon abundance from the
line in the RCB star spectra.
The transition probability for the 9850 \AA\ line is taken from
Galavis, Mendoza \& Zeippen (1997) who give A = 2.23 $\times 10^{-4}$ s$^{-1}$ or
$\log gf = -10.79$ with an uncertainty of about 0.04 dex.   
The lower excitation potential is a mere 0.005 eV.

Two additional lines complete multiplet 1F.
The stronger line with a $gf$-value about 3 times smaller than for the 9850 \AA\ line is
at 9824.30 $\pm$ 0.05 \AA\ (Liu et al. 1995) and
irretrievably blended in the spectra of cooler RCB stars (GU\,Sgr, V482\,Cyg, SU\,Tau, and R\,CrB).
%, being sandwiched between N\,{\sc i} and O\,{\sc i} lines
In the spectra of VZ\,Sgr, UV\,Cas, and XX\,Cam, 
the 9824.30 \AA\ line is absent, but this is as expected from the S/N and
the estimated strength of the observed 9850 \AA\ line.
The third [C\,{\sc i}] line at 9808.3 \AA\ is expected to be 4000
times weaker than the 9850 \AA\ line, and  undetectable.
(The multiplet 2F $2p^2$ $^3$P$_{1}$ - $2p^2$ $^1$S$_{0}$ transition
at 4621.57 \AA\ with $\log gf$ = $-$11.12 is almost certainly blended as the
blue spectral region of an RCB star is rich in strong lines.)

\subsection{Overview of the Spectra}

The region around 9850 \AA\ is shown in Figure 1 where the
stars are ordered by increasing temperature from bottom to top.
The spectrum of the  normal yellow supergiant $\gamma$ Cyg is shown at the
bottom. In the spectra of the hottest stars, the strongest
lines are from C\,{\sc i} and N\,{\sc i} transitions with weaker
lines of Si\,{\sc i}, Fe\,{\sc i}, and Fe\,{\sc ii}. The [C\,{\sc i}]
line falls in the red wing of a Fe\,{\sc ii} line. Spectra of the
coolest stars GU\,Sgr and V482\,Cyg show additional lines which we identify as 
high rotational lines of the 1-0 band of the CN molecule's Red System
(Davis \& Phillips 1963). These CN lines are prominent in the
spectra of GU\,Sgr and V482\,Cyg and traceable in SU\,Tau, R\,CrB and
RY\,Sgr. Two CN lines bracket the [C\,{\sc i}] line.

Inspection of Figure 1 shows several additional features of
interest. As anticipated, the
strength of a given C\,{\sc i} line is unchanged along the
temperature sequence from XX\,Cam at the high temperature
end to GU\,Sgr at the low temperature end. Also anticipated
is the very large width of all absorption lines in the
spectra of RCB stars. Compare the widths with those of lines in $\gamma$ Cyg,
which itself is commonly referred to as showing broad lines.
A surprise  is
the appearance of emission in the RY\,Sgr spectrum near the
wavelength of the [C\,{\sc i}] line. The illustrated spectrum is from
31 July 2002 in which the 8727 \AA\ line is also in
emission. A spectrum from 22 June 1997 also shows emission
but to the red of the wavelength of the [C\,{\sc i}] line.
The possibility of emission contaminating the
absorption line is an unfortunate complication.

\begin{figure}
\epsfxsize=8truecm
\epsffile{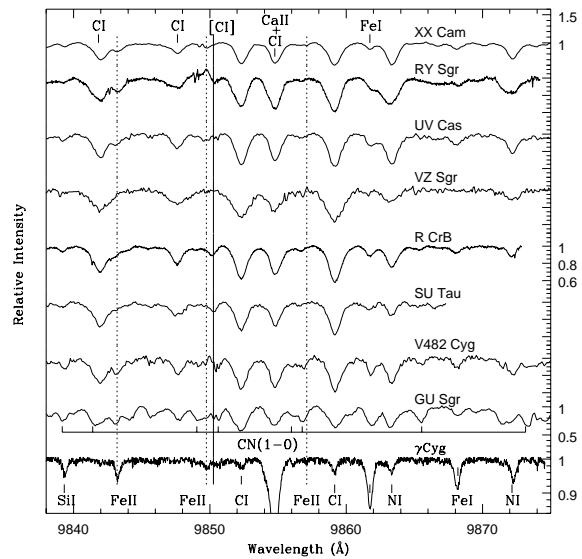}
\caption{The spectra of the RCB stars in the region of [C\,{\sc i}] 9850.264\AA\
line, which is indicated by a vertical solid line, are plotted with their effective temperatures 
increasing from bottom to
top. The spectrum of $\gamma$ Cyg is also plotted. The positions of
a number of other atomic and some CN lines are also
marked and the vertical dashed lines indicate the Fe\,{\sc ii} features. The relative
intensity scale for GU\,Sgr and XX\,Cam is different from the other RCB stars.}
\end{figure}

\begin{figure}
\epsfxsize=8truecm
\epsffile{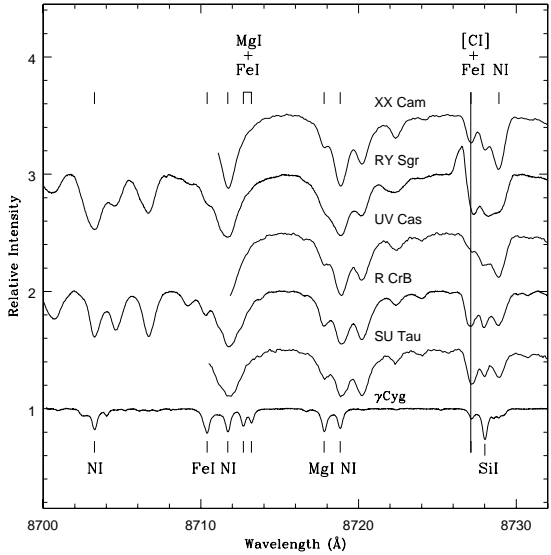}
\caption{The spectra of RCB stars in the region of [C\,{\sc i}] 8727.126\AA\
line, which is indicated by a vertical solid line, are plotted with their effective temperatures
increasing from bottom to top.
The spectrum of $\gamma$ Cyg is also plotted. The positions of the key lines
are also marked.}
\end{figure}

%\subsection{The  8727 \AA\ Region}

%The remarks made about the 9850 \AA\ region apply in large measure to
%the region around the 8727 \AA\ [C\,{\sc i}] line, which  is
%stronger than the 9850 \AA\ line 
% (Figure 2).
 The [C\,{\sc i}] 8727 \AA\ line (Figure 2)
is slightly blended with a Si\,{\sc i} line. Emission
is very pronounced in the spectrum of RY\,Sgr. The central line depths
of the [C\,{\sc i}] line run from about 16 \% for SU\,Tau to
36 \% for RY\,Sgr with no obvious systematic trend with the
effective temperature.

\section{Spectrum Synthesis - Procedure}

The fact that the [C\,{\sc i}] lines are blended with other
lines requires that
the carbon abundance be extracted  by
spectrum synthesis. 
%This technique requires a
%reliable model atmosphere and atomic data for the offending
%blends. 
In addition to synthesizing the [C\,{\sc i}] line and its
blends, we use lines of C\,{\sc i}, N\,{\sc i}, Si\,{\sc i},
Fe\,{\sc i} and Fe\,{\sc ii} to obtain abundances and compare with
values given by Asplund et al. (2000). 
%from their analysis of lines
%at shorter wavelengths.

Synthetic spectra are generated using the H-deficient model
atmospheres computed by Asplund et al. (1997) and the Uppsala
spectrum synthesis code BSYNRUN. Models corresponding to a C/He ratio of
1 \% by number were chosen. This composition
is equivalent to a carbon abundance $\log\epsilon$(C) = 9.54
on the scale $\Sigma\mu_i\epsilon_i = 12.15$, where
$\mu_i$ is the mean atomic weight of element $i$. Stellar parameters
derived by Asplund et al. (2000) 
%in their comprehensive abundance analysis 
were adopted as our initial values. Synthetic
spectra  were convolved
with a Gaussian profile to account for the combined broadening of
stellar lines by the atmospheric macroturbulent velocity, rotational
broadening, and the instrumental profile.

\subsection{The 9850 \AA\ Region}

The 9850 \AA\ [C\,{\sc i}] line
may be blended with Fe\,{\sc ii}  and CN lines. The Fe\,{\sc ii}
line  at 9849.74 \AA\ has a lower excitation potential of 6.729 eV.
To obtain the $gf$-value of the line, we use the spectrum of $\gamma$
Cyg where the Fe\,{\sc ii} line is present. With a model
atmosphere constructed for the stellar parameters and iron abundance
given by Luck \& Lambert (1981), we find $\log gf$ = $-$2.60. The
basic data for the CN 1-0 lines were taken from Kurucz's
line list and
checked against the SCAN tape (J{\o}rgensen \& Larsson 1990)
and a recipe given by Bakker \& Lambert (1998). A dissociation
energy of  7.65 eV was adopted for the molecule (Bauschlicher, Langhoff \&  Taylor
1988). Unblended CN lines away from the 9850 \AA\ feature
were fitted by adjusting the N abundance, and this abundance
used in computing the CN lines blended with the [C\,{\sc i}]
line.
An estimate of the N abundance was obtained not only from the
CN lines but also from N\,{\sc i} lines near the [C\,{\sc i}] lines.
The $gf$-values for the N\,{\sc i} lines were taken from 
Wiese, Fuhr \& Deters (1996) compilation. 

%For a direct comparison of the [C\,{\sc i}] line with the C\,{\sc i}
%lines, we include two C\,{\sc i} lines in the synthetic spectrum
%around 9850 \AA. 
The C\,{\sc i} lines at 9852.27 \AA,  a
blend of two lines, and 9859.15 \AA\
(Figure 1) have also been used for estimating the C abundance. 
Their $gf$-values are determined by inversion of their
equivalent widths  in the $\gamma$ Cyg spectrum using 
%the model previously mentioned and 
the C abundance derived
from a set of weak C\,{\sc i} lines at shorter wavelengths. 
%and  the
%8727 \AA\ [C\,{\sc i}] line which is clearly seen in
%Figure 2. The 9850 \AA\ [C\,{\sc i}] line is not detected in
%our spectrum of $\gamma$ Cyg.

\subsection{The 8727 \AA\ Region}

The red wing of the  [C\,{\sc i}] line is blended with a Si\,{\sc i}
line (Figure 2). The $gf$-value of the latter line is
derived from its equivalent width in the solar spectrum. The
$gf$-value of the N\,{\sc i} line to the red of the Si\,{\sc i}
line is taken from Wiese, Fuhr \& Deters (1996). Spectrum synthesis of
this region was confined to a 6 \AA\ window around the
[C\,{\sc i}] line.

\section{Spectrum Synthesis - Results}

%In the Introduction, we noted the `carbon problem'
%(Asplund et al. 2000). The C\,{\sc i} lines in RCB spectra are
%much weaker than predicted. Quantitatively, the effect corresponds to
%a factor of four reduction in the product of a line's $gf$-value and
%the C abundance. 
The [C\,{\sc i}] lines were sought in order to
shed new light on the `carbon problem'. Asplund et al. (2000)
suggested that the 9850 \AA\ [C\,{\sc i}]
line was not subject to the carbon problem. Their analysis was
flawed in that the contributions of the blending Fe\,{\sc ii}
and CN line were not considered.

In addition to fitting the [C\,{\sc i}] lines, we analysed other lines
in order to check the abundances given by Asplund
et al. (2000). To within the uncertainties of the analyses and with the
exception of V482\,Cyg (see below), we confirm the result for
carbon using C\,{\sc i} lines in the 9850 \AA\ region, the
N abundance using N\,{\sc i} lines in the 9850 \AA\ and 8727 \AA\ regions,
the Si abundance from Si\,{\sc i} lines in the  9850 \AA\ and 8727 \AA\ regions, and the Fe
abundance from Fe\,{\sc i} and Fe\,{\sc ii} lines in the 9850 \AA\ region.
The CN 1-0 lines are also used to check the N abundance.
In these atmospheres, molecules are trace species whose formation
does not affect the partial pressures of the constituent
atoms. Then,
consideration of the ratio of line to continuous opacity shows that 
the strength of a CN line should be dependent on the N abundance
 but  independent
of the C abundance.

  Following remarks about $\gamma$ Cyg,
 we discuss the RCB
stars in order of decreasing effective temperature.
It should be noted that the C abundances given here for the RCB
stars are not self-consistent -- the model atmospheres and their 
continuous absorption were calculated assuming a logarithmic
C abundance of 9.54. If the same abundance is adopted
for the calculations of the  carbon lines -- permitted and forbidden --
the predicted  and observed equivalent widths
are in disagreement. Thus, the C abundance derived here for the
RCB stars is basically a measure of the inconsistency, which we
term the carbon problem. 

\begin{figure}
\epsfxsize=8truecm
\epsffile{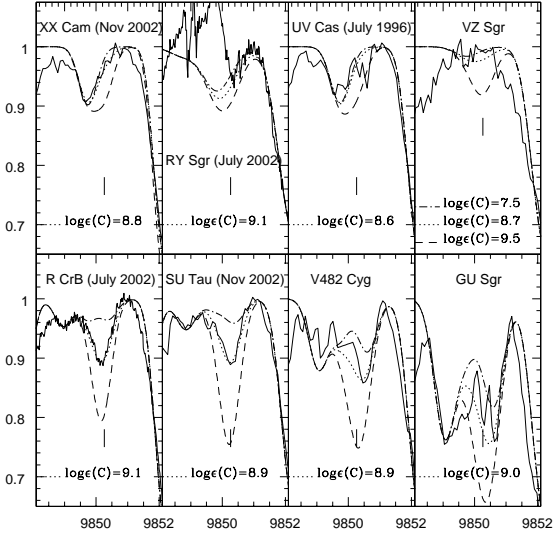}
\caption{Observed (solid line) and synthetic
[C\,{\sc i}] 9850.264\AA\ line profiles of RCB stars.
Synthetic profiles, including the blends, are shown for each star for three different
abundances. The maximum abundance of $\log\epsilon$(C) = 9.5, the model
input abundance is shown by the long dashes and in no case does this fit the
observed feature. The minimum abundance of $\log\epsilon$(C) = 7.5 shows in
effect the predicted spectrum for no contribution from the [C\,{\sc i}]
line (see dot-dash line). The best-fitting synthetic spectrum
is shown by the line made of short dashes and the corresponding
abundance is indicated in each panel.}
\end{figure}

\subsection{$\gamma$ Cyg}

Our analysis of permitted and forbidden C\,{\sc i} lines uses
a MARCS model atmosphere (Gustafsson et al. 1975)
for the atmospheric parameters derived by Luck \& Lambert  (1981).
The 8727 \AA\ [C\,{\sc i}] line (Figure 2) yields the abundance
$\log\epsilon$(C) = 7.88.
The 9850 \AA\ [{C\,{\sc i}] line is not detectable and the upper limit to
its equivalent width corresponds to $\log\epsilon$(C) $\leq$ 8.18.
 A selection of 14 C\,{\sc i} lines from 5380 \AA\ to 8873 \AA\ with
$gf$-values from Wiese, Fuhr \& Deters (1996) give the abundance
$\log\epsilon$(C) = 7.94 $\pm$ 0.14. These lines range
in equivalent width from 6 m\AA\ to 82 m\AA. 
%The carbon abundance is sub-solar because the first dredge-up has
%brought CN-cycled material into the atmosphere. The N abundance
%is correspondingly supra-solar (Luck \& Lambert 1981).
For $\gamma$ Cyg, it is seen that the C\,{\sc i} lines and the
8727 \AA\ forbidden line return the same abundance within the uncertainties.
This result is in sharp contrast to the  results
from the RCB stars.

%The C\,{\sc i} lines in the 9850 \AA\ region cannot be used 
%in the abundance analysis because
%their $gf$-values have not been determined by either theoretical or
%experimental means. These lines, however, are used to provide an
%`astrophysical' estimate of the $gf$-value for lines used in the
%construction of synthetic spectra for the RCB stars. 

\subsection{XX Cam}

 The 9850 \AA\ feature is dominated by
the Fe\,{\sc ii} line such that we can set only an
upper limit to the [C\,{\sc i}] contribution and the associated
C abundance of $\log\epsilon$(C) $\leq$ 8.8 (Figure 3). 
The 8727 \AA\ line (Figure 4) is fit with an abundance
$\log\epsilon$(C) = 8.4.  The limit from the 9850 \AA\ line
to the C abundance is consistent with that from the 8727 \AA\ line.
The two C\,{\sc i}
lines near 9850 \AA\ give an abundance
consistent with the C abundance of $\log\epsilon$(C) = 9.0 $\pm$ 0.4
from Asplund et al. who used a collection of C\,{\sc i} lines and 
Opacity Project $gf$-values (Lou \& Pradhan 1989; Hibbert et al. 1993;
Seaton et al. 1994).

Asplund et al. put  the uncertainty in $T_{\rm eff}$  at about
$\pm$ 250 K which corresponds to 
corrections to abundances derived from   C\,{\sc i} and
[C\,{\sc i}] lines of  0.05, and 0.2 dex, respectively.
The uncertainty in $\log g$ $\simeq \pm$ 0.5 provides only
minor corrections to the abundances. An additional
potential source of uncertainty is the adopted microturbulent
velocity of $\xi = 9.0$ km s$^{-1}$ (Asplund et al. 2000). A change of
$\xi$ by $\pm$ 2 km s$^{-1}$ changes the C  abundance by 0.2 dex
from the C\,{\sc i} lines and by only 0.05 dex from the [C\,{\sc i}]
8727 \AA\ line. These estimates of uncertainty do not allow the 0.6 dex
difference in the C\,{\sc i} and [C\,{\sc i}] abundances. 
%to be
%attributed to uncertainties arising from the adopted atmospheric
%parameters. This conclusion holds for the other stars
%where the 8727 \AA\ is observed.

%The original carbon problem uncovered by Asplund et al. is confirmed,
%the C\,{\sc i} lines give an abundance 0.5 dex less than
%the input abundance of the adopted model. 
The carbon problem for the
[C\,{\sc i}] line is even more severe as seen for
8727 \AA\ [C\,{\sc i}] line which gives an abundance 0.6 less than the
C\,{\sc i} lines and 1.1 dex less than the input abundance.

\begin{figure}
\epsfxsize=8truecm
\epsffile{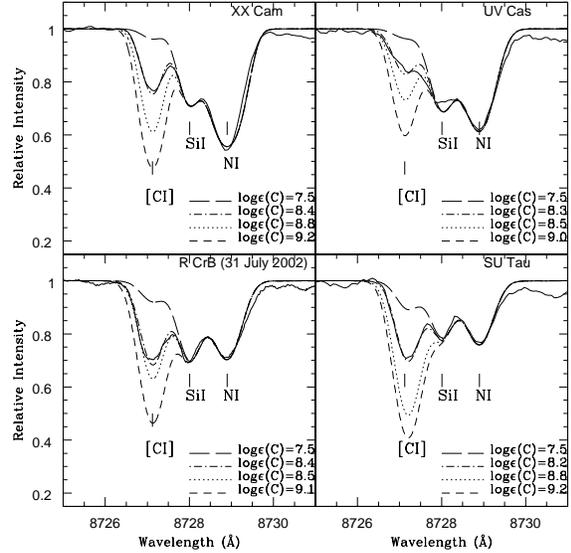}
\caption{Observed [C\,{\sc i}] 8727.126\AA\ line profiles (solid lines) of
RCB stars. Synthetic spectra are shown for four C abundances, as shown on the figure.}
\end{figure}

\subsection{RY Sgr}

The 31 July 2002 spectrum  shows
emission in the blue wing of the 8727 \AA\ forbidden
line.  Peak emission occurs  at the velocity of $-$29 km s$^{-1}$.
This is
blue-shifted  by about 17 km s$^{-1}$ from the photospheric
velocity of $-$12 km s$^{-1}$. The emission line has a width (FWHM) of about
19 km s$^{-1}$. A spectrum  taken the same
night shows emission at 9850 \AA\ (Figure 1).
Emission at [C\,{\sc i}] may be a common occurrence for this star.
Our 1997 June 22 spectrum shows 9850 \AA\ in emission at the
radial velocity of $-$10.5 km s$^{-1}$ equivalent to a red-shift of
19.5 km s$^{-1}$ relative to the photosphere.
This spectrum does not
include the region around the 8727 \AA\ line.

RY\,Sgr exhibits a pulsation with an amplitude of about 35 km s$^{-1}$
(Lawson, Cottrell \& Clark 1991). The mean or systemic velocity was given as $-$21 km s$^{-1}$ by
Lawson, Cottrell \& Clark (1991). Relative to this velocity the emission at 8727 \AA\
in the 2002 spectrum is blue-shifted by about 8 km s$^{-1}$. 
Perhaps, the emission at  maximum light  is related to the
fact that RY\,Sgr is a  large-amplitude radial velocity
pulsator.  
%At certain phases the absorption lines
%are split into two components (Cottrell \& Lambert 1982)
%suggesting passage of a shock through the atmosphere. This
%shock might  be the site of emission. 
There is no hint of emission in our
spectra of the other RCB stars, even in V482\,Cyg observed 
below maximum light.

By inspection, the 9850 \AA\ blend of Fe\,{\sc ii} and [C\,{\sc i}]
lines is very similar to that of XX\,Cam and, hence, the forbidden
line is a minor contributor to the absorption feature.
Synthesis of the 8727 \AA\ line 
suggests that emission overlays the absorption line and one might
suppose that an abundance $\log\epsilon$(C) $\sim$ 9.2 is a lower
limit but this  assumes that the emission
comes from a region exterior to the regions in which the absorption
line is formed. This is not necessarily a valid assumption 
for RY\,Sgr with its large pulsation and evidence for an internal
shock. We decline to quote a carbon abundance. But, a clue might be
derived from RY\,Sgr observations that the emission effects both [C\,{\sc i}]
lines. The estimations from the red absorption wing (assumed to be 
unaffected by emissions) of both these [C\,{\sc i}] lines suggests the 
same carbon abundance of about 9.2 which agrees with that from 
C\,{\sc i} lines.

%Inspection of Figure 1 shows that the CN lines are present. 
With the
adopted model, the CN lines return a N abundance about 0.6 dex higher than
from the N\,{\sc i} lines used here and by Asplund et al. This 
difference between N abundances can be eliminated by lowering the adopted
$T_{\rm eff}$ by about 200 K, or by a combination of a smaller $T_{\rm eff}$
change accompanied by a reduction of $\log g$. These adjustments are
within the uncertainty limits considered reasonable  by Asplund et al. 

%\begin{figure}
%\epsfxsize=8truecm
%\epsffile{fig5.ps}
%\caption{Observed [C\,{\sc i}] 8727.126\AA\ line profile of RY\,Sgr (solid lines).
% Synthetic spectra are shown for four C abundances, as shown on the figure.}
%\end{figure}

\subsection{UV Cas}

The 8727 \AA\ line is reasonably well fit with the abundance
$\log\epsilon$(C) = 8.3 (Figure 4). Two observations 
show no convincing evidence for the 9850 \AA\ [C\,{\sc i}]
line, indicating $\log\epsilon$(C) $\leq$ 8.7 (Figure 3). The two
C\,{\sc i} lines in the 9850 \AA\ region suggest an abundance close to
$\log\epsilon$ = 9.0, a value only 0.2 dex less than that obtained by
Asplund et al. (2000). The CN lines are not detectable in our spectra.
UV\,Cas joins XX\,Cam in showing a larger carbon
problem for the 8727 \AA\ line than for the C\,{\sc i} lines.

%\begin{figure}
%\epsfxsize=8truecm
%\epsffile{fig6.ps}
%\caption{Observed [C\,{\sc i}] 8727.126\AA\ line profile of UV\,Cas (solid lines).
% Synthetic spectra are shown for four C abundances, as shown on the figure.}
%\end{figure}

\subsection{VZ Sgr}

This RCB star was classified as a `minority' star by Lambert \&
Rao (1994), i.e., the metal lines are weak.
 This is
evident from Figure 1 which shows that the Fe\,{\sc i} and Fe\,{\sc ii} 
lines are absent from VZ\,Sgr's spectrum. Unfortunately,  the region around
8727 \AA\ has not yet been observed. 
%The 9850 \AA\ forbidden carbon line is not detected. 
Synthesis (Figure 3) suggest an upper limit
$\log\epsilon$(C) $\leq$ 8.7 for the [C\,{\sc i}] 9850 \AA\  line, a limit clearly below the input
abundance of 9.5. The two C\,{\sc i} lines give 
$\log\epsilon$(C) = 9.1. Asplund et al. gave $\log\epsilon$(C) = 
8.8 $\pm$ 0.3.

\subsection{R CrB}

At 8727 \AA\, the synthesis corresponding to $\log\epsilon$(C) $\simeq$ 8.4
provides a fair fit to the observed high-resolution profile of 2002 July 31 (Figure 4).
A lower resolution (R = 60000) spectrum of 1995 June 17 provides
the same abundance. In neither spectra is there  a hint that
emission has distorted the absorption profile.

The 9850 \AA\ syntheses  include
the CN lines blending with the [C\,{\sc i}] and Fe\,{\sc ii}
feature. On the high-resolution spectrum of
2002 July 31, 
the CN 1-0 lines are blue-shifted by  3.6 km s$^{-1}$ relative to the
high-excitation lines of C\,{\sc i} and N\,{\sc i}.
Since the [C\,{\sc i}] line  and CN lines are all of low
excitation, we assume  
that they have a common  blueshift.
 Synthetic spectra (Figure 3) indicate
an abundance $\log\epsilon$(C) = 9.1. Examination of our library
of 9850 \AA\ spectra of R\,CrB (e.g., Rao \& Lambert 1997; Rao et al.
1999) shows that the velocity shift between CN and the C\,{\sc i} and
N\,{\sc i} varies during the star's pulsation and this shift
ranges from $-$3.0 to $+$4.0 km s$^{-1}$.
Syntheses taking into the account the velocity shift return the
same abundance from four different spectra.
If the blue-shift is neglected, the fit to the 9850 \AA\ feature
is less satisfactory but the derived carbon abundance is little
affected.
Asplund et al.'s atmospheric parameters
 are used in all cases.
%, even though Rao \& Lambert suggested these parameters change slightly during the pulsation. 
The C\,{\sc i} lines near 9850 \AA\ are well fitted with a
similar abundance ($\log\epsilon$(C) = 9.2), an abundance  equal to that
determined by Asplund et al. 

%As in the case of XX\,Cam,  the [C\,{\sc i}] lines add a new
%dimension to the carbon problem,
The 9850 \AA\ line confirms the
abundance obtained from C\,{\sc i} lines, but the 8727 \AA\
line gives an abundance 0.7 dex below that from the 9850 \AA\ line or
1.1 dex below the input carbon abundance of the model.

The N\,{\sc i} lines in the 8727 \AA\ and 9850 \AA\ regions give
the abundance $\log\epsilon$(N) = 8.2, a value consistent
with the result of 8.4 $\pm$ 0.2 given by Asplund et al. 
The observed CN 1-0
lines are well matched with the abundance $\log\epsilon$(N) = 8.2.
This may be fortuitous agreement because the sensitivity of the
CN line strengths to a change of the atmospheric parameters,
especially to $T_{\rm eff}$, is high. Note that a change of 250 K changes
the required abundance by 0.5 dex.

%\begin{figure}
%\epsfxsize=8truecm
%\epsffile{fig7.ps}
%\caption{Observed [C\,{\sc i}] 8727.126\AA\ line profile of R\,CrB (solid lines).
% Synthetic spectra are shown for four C abundances, as shown on the figure.}
%\end{figure}

\subsection{SU Tau}

%Synthesis of the 15 November 2002 spectrum shows that 
The abundance $\log\epsilon$(C) = 8.2
provides an excellent fit to the observed [C\,{\sc i}] line
at 8727 \AA\ (Figure 4),
%The 9850 \AA\ feature is dominated
%by the contribution from the [C\,{\sc i}] line. 
whereas $\log\epsilon$(C) = 8.9 is required for an equivalent fit 
to the observed 9850 \AA\ feature (Figure 3).
For this fit, the  CN lines are blue-shifted by 6 km s$^{-1}$ from the
high excitation atomic lines. This shift is assumed to apply to the
[C\,{\sc i}] line also, but the derived abundance is not
critically influenced by the shift. The abundance from the 9850 \AA\ 
[C\,{\sc i}] line is consistent with that derived by us from the
9850 \AA\ region's C\,{\sc i} lines and the C\,{\sc i} lines
used by Asplund et al. In contrast and consistent with the results for
XX\,Cam and R\,CrB, the 8727 \AA\ line  gives a markedly
lower carbon abundance.  

The N abundance reported by Asplund et al. was based on just
two
N\,{\sc i} lines. Examination of our superior spectra provides a
more accurate equivalent width for one line. The other line is
not present on our spectrum. We identify a third line.
We adopt
Wiese et al.'s $gf$-value and derive the N abundance
$\log\epsilon$(N) = 7.9. We suggest that this is a more reliable
estimate than Asplund et al.'s value of 8.5. 
The N abundance estimate from the CN lines is 0.5 dex lower.
 The difference is erasable with
minor adjustments to the adopted atmospheric parameters, such as
an increase of $T_{\rm eff}$ by only 140 K.

%\begin{figure}
%\epsfxsize=8truecm
%\epsffile{fig8.ps}
%\caption{Observed [C\,{\sc i}] 8727.126\AA\ line profile of SU\,Tau (solid lines).
% Synthetic spectra are shown for four C abundances, as shown on the figure.}
%\end{figure}

\subsection{V482 Cyg}

%Our spectrum includes the 9850 \AA\ but not the 8727 \AA\ region. 
There is a hint of emission at the 9850 \AA\
[C\,{\sc i}] line. Examination of the three
individual exposures shows that this
`emission' occurs  in only one exposure that being the one with the 
lowest signal-to-noise ratio. We, therefore, consider the emission
to be an artefact which is removed in Figure 3. A fit to the absorption 
feature suggests an abundance upper limit $\log\epsilon$(C) $\leq$ 8.9.

In conflict with this upper limit,
the C abundance of  $\log\epsilon$(C) = 9.5$\pm$0.3 is derived
from C\,{\sc i} lines in our spectrum. Asplund et al. 
obtained $\log\epsilon$(C) = 8.9 from C\,{\sc i} lines.
Examination of our spectrum shows that C\,{\sc i} lines
across the spectrum are systematically stronger than reported
by Asplund et al. This strengthening and the adoption of the
same model as Asplund et al. necessarily results in a higher carbon
abundance. Recall that the star was two magnitudes below maximum
light. We suppose that the atmospheric structure differed from
that prevailing at maximum light. The temperature gradient
may have been steeper in decline than at maximum light. A very
similar strengthening of C\,{\sc i} (and other) lines was
noted by Rao et al. (1999) in spectra of R\,CrB taken early in 
a deep decline.

The 
%9850 \AA\ 
N\,{\sc i} lines give an abundance 0.8 dex
higher than that from the CN lines. 
Erasure of the N\,{\sc i} -- CN discrepancy requires
a different choice of atmospheric parameters. The possibilities
range from raising $T_{\rm eff}$ by 400 K at the adopted
$\log g$ to lowering $\log g$ by about 0.8 dex at the adopted
$T_{\rm eff}$. These are not unacceptable changes given that
the atmosphere may have been perturbed.

\subsection{GU Sgr}

%Our observations cover the 9850 \AA\ but not the 8727 \AA\ region.  
The blend containing the 9850 \AA\ [C\,{\sc i}] line
is contaminated with CN lines which are strong in this spectrum.
Our syntheses
suggests an upper limit $\log\epsilon$(C) $\leq$ 9.0,
a value consistent with Asplund et al.'s value of 8.8.

Analysis of the CN lines gives a N abundance of
$\log\epsilon$(N) = 8.1. Since the N\,{\sc i} lines in the
9850 \AA\ region are blended, we compare the CN-based abundance
with the result $\log\epsilon$(N) = 8.7 $\pm$ 0.5 given by
Asplund et al. The difference may be removed by a modest (250 K)
increase of the adopted $T_{\rm eff}$ at the adopted $\log g$.

\subsection{Summary of the Carbon Abundances}

%\begin{table*}
%\centering
%\begin{minipage}{105mm}
%\caption{Photospheric C abundances from C\,{\sc i} and [C\,{\sc i}] lines}
%\begin{tabular}{lccccccc} \hline
%\multicolumn{1}{c}{}& \multicolumn{3}{c}{Model$^a$}&  \multicolumn{1}{c}{}&\multicolumn{3}{c}{log$\epsilon$(C)} \\
%\cline{2-4} \cline{6-8}
%\multicolumn{1}{c}{Star}& \multicolumn{1}{c}{$T_{\rm eff}$} & \multicolumn{1}{c}{log$g$} & \multicolumn{1}{c}{$\xi$}&
%\multicolumn{1}{c}{}&\multicolumn{1}{c}{C\,{\sc i}$^a$} &\multicolumn{1}{c}{$[\rm C\,{\sc i}]$} &\multicolumn{1}{c}{$[\rm C\,{\sc i}]$}\\
%\multicolumn{1}{c}{}& \multicolumn{1}{c}{K}& \multicolumn{1}{c}{cm s$^{-2}$}&\multicolumn{1}{c}{km s$^{-1}$}&\multicolumn{1}{c}{}&
%\multicolumn{1}{c}{}&\multicolumn{1}{c}{9850\AA}&\multicolumn{1}{c}{8727\AA}\\ \hline
%XX~Cam & 7250 & 0.75 & 9.0 & & 9.0 & $<$8.8 & 8.4\\
%RY~Sgr & 7250 & 0.75 & 6.0 & & 8.9 & ... & ...    \\
%UV~Cas & 7250 & 0.50 & 7.0 & & 9.2 & $<$8.7 & 8.3 \\
%VZ~Sgr & 7000 & 0.50 & 8.0 & & 9.1 & $<$8.7 & ... \\
%R~CrB  & 6750 & 0.50 & 7.0 & & 9.2 & 9.1 & 8.4 \\
%SU~Tau & 6500 & 0.50 & 7.0 & & 8.8 & 8.9 & 8.2 \\
%V482~Cyg& 6500 & 0.50 &4.0 & & 8.9 & $<$8.9 & ...\\
%GU~Sgr & 6250 & 0.50 & 7.0 & & 8.8 &$<$9.0 & ... \\
%$\gamma$ Cyg$^b$ & 6100 & 0.55 & 3.5 & & 7.9 & $<$8.2 & 7.9 \\
%\hline
%\end{tabular}
%$^a$Asplund et al. (2000) for R~CrB stars; Luck \& Lambert (1981) for $\gamma$ Cyg\\
%$^b$see Section 5.1 of the text
%%$^a$covers 8727\AA\ region\\
%%$^b$covers the 8727\AA\ and 9850\AA\ regions
%\end{minipage}
%\end{table*}
\begin{table*}
\centering
\begin{minipage}{105mm}
\caption{Photospheric C abundances from C\,{\sc i} and [C\,{\sc i}] lines}
\begin{tabular}{lcccccccc} \hline
\multicolumn{1}{c}{}& \multicolumn{3}{c}{Model$^a$}&  \multicolumn{1}{c}{}&\multicolumn{4}{c}{log$\epsilon$(C)} \\ 
\cline{2-4} \cline{6-9}
\multicolumn{1}{c}{Star}& \multicolumn{1}{c}{$T_{\rm eff}$} & \multicolumn{1}{c}{log$g$} & \multicolumn{1}{c}{$\xi$}&
\multicolumn{1}{c}{}&\multicolumn{1}{c}{C\,{\sc i}$^a$} &\multicolumn{1}{c}{C\,{\sc i}$^b$} &\multicolumn{1}{c}{$[\rm C\,{\sc i}]$} 
&\multicolumn{1}{c}{$[\rm C\,{\sc i}]$}\\
\multicolumn{1}{c}{}& \multicolumn{1}{c}{K}& \multicolumn{1}{c}{cm s$^{-2}$}&\multicolumn{1}{c}{km s$^{-1}$}&\multicolumn{1}{c}{}&
\multicolumn{1}{c}{} &\multicolumn{1}{c}{} &\multicolumn{1}{c}{9850\AA}&\multicolumn{1}{c}{8727\AA}\\ \hline
RCB stars: &  &      &     & &     &     &        & \\
XX~Cam & 7250 & 0.75 & 9.0 & & 9.0 & 8.9 & $<$8.8 & 8.4\\
RY~Sgr & 7250 & 0.75 & 6.0 & & 8.9 & 9.3 & ... & ...    \\
UV~Cas & 7250 & 0.50 & 7.0 & & 9.2 & 9.0 & $<$8.7 & 8.3 \\
VZ~Sgr & 7000 & 0.50 & 8.0 & & 9.1 & 9.1 & $<$8.7 & ... \\
R~CrB  & 6750 & 0.50 & 7.0 & & 9.2 & 9.2 & 9.1 & 8.4 \\
SU~Tau & 6500 & 0.50 & 7.0 & & 8.8 & 9.0 & 8.9 & 8.2 \\
V482~Cyg& 6500 & 0.50 &4.0 & & 8.9 & 9.6 & $<$8.9 & ...\\
GU~Sgr & 6250 & 0.50 & 7.0 & & 8.8 & 9.3 & $<$9.0 & ... \\
Standard star: &  &      &     & &     &     &        & \\
$\gamma$ Cyg$^c$ & 6100 & 0.55 & 3.5 & & 7.9 & ... & $<$8.2 & 7.9 \\
\hline
\end{tabular}
$^a$Asplund et al. (2000) for R~CrB stars, the expected C abundance is 9.54 when C/He of 1\% models are used; Luck \& Lambert (1981) for $\gamma$ Cyg\\
$^b$based on two C\,{\sc i} lines near 9850\AA\\
$^c$see Section 5.1 of the text
%$^a$covers 8727\AA\ region\\
%$^b$covers the 8727\AA\ and 9850\AA\ regions
\end{minipage}
\end{table*}

Our analyses use the appropriate MARCS model for the
atmospheric parameters recommended by Asplund et al. 
and C/He = 1\% by number. This input
abundance corresponds to a carbon abundance $\log\epsilon$(C) = 9.5.
In Table 2, we summarize the carbon abundances derived from
the two [C\,{\sc i}] lines,  
and the C\,{\sc i}-based abundance given by
Asplund et al. illustrating the carbon problem.
%Across the table there is a carbon problem, in that the tabulated
%abundances are systematically less than this input abundance.

Except for V482\,Cyg, the carbon abundance derived from
the two 9850 \AA\ region  C\,{\sc i} lines
 is in good agreement with that obtained
by Asplund et al. 
%from a larger sample of C\,{\sc i} lines at
%shorter wavelengths. 
%Recall that V482 Cyg was observed below
%maximum light.

%Detection of the 9850 \AA\ forbidden line for R\,CrB and SU\,Tau
%leads to a carbon abundance consistent with that from the
%C\,{\sc i} lines.
%Even for stars where 
%an upper limit to the carbon abundance is provided from the forbidden
%line  at
%9850 \AA\, the limit is clearly below the input abundance 
Table 2 shows 
that a carbon problem extends to the 9850 \AA\ forbidden line.
This conclusion is contrary to that reached by Asplund et al.
from analysis of equivalent widths of the 9850 \AA\ line.
The explanation
is that Asplund et al. did not recognize 
%in their lower resolution spectra 
that the [C\,{\sc i}] line was a
blend. In the case of V482\,Cyg, which was observed two magnitudes below
maximum light, the carbon problem has vanished for the C\,{\sc i}
lines but not for the 9850 \AA\ [C\,{\sc i}] line. This is the
only case where the C\,{\sc i} lines and 9850 \AA\ differ in the
abundance they provide. 

Surprisingly, the 8727 \AA\ [C\,{\sc i}] line offers further
information on
the carbon problem. For each of the four stars for
which we have observed the 8727 \AA\  line, the derived
abundance is less than that from the C\,{\sc i} lines. 
In the case of R\,CrB and SU\,Tau, 
%the two stars for which a carbon
%abundance is obtained from both forbidden lines,
 the 8727 \AA\ line gives an abundance 0.7 dex
less than that derived from the 9850 \AA\  line. 
That the two forbidden lines give very different abundances is
especially puzzling if these lines are formed, as expected, in or close
to LTE and, as might be supposed, are a product of  the stellar
photosphere, i.e., a region with the temperature decreasing
monotonically outward. 
An unidentified atomic line superimposed on the 8727 \AA\
forbidden line would mean that we have overestimated
the carbon abundance from that line. 
In this scenario, the carbon problem is more severe
for both of the [C\,{\sc i}] lines than for the C\,{\sc i} lines.
However, it is difficult to find a carrier for a line which
is strong in R\,CrB and SU\,Tau but weak in other RCB stars and
$\gamma$ Cyg.

%\begin{table}
%\centering
%\begin{minipage}{55mm}
%\caption{Carbon abundances for R\,CrB for various model atmospheres}
%\begin{tabular}{cccccc} \hline
%\multicolumn{1}{c}{}&\multicolumn{1}{c}{}&\multicolumn{1}{c}{}& \multicolumn{3}{c}{log$\epsilon$(C)} \\
%\multicolumn{3}{c}{Model}& \multicolumn{1}{c}{C\,{\sc i}} &
%\multicolumn{1}{c}{$[\rm C\,{\sc i}]$} &\multicolumn{1}{c}{$[\rm C\,{\sc i}]$}\\
%\cline{1-3}
%\multicolumn{3}{c}{$T_{\rm eff}$,log$g$,$\xi$}& \multicolumn{1}{c}{}&
%\multicolumn{1}{c}{9850\AA}&\multicolumn{1}{c}{8727\AA}\\ \hline
%&6750,0.5,7&&9.2&9.1&8.4\\
%&6500,0.5,7&&9.2&8.9&8.2\\
%&7000,0.5,7&&9.3&9.3&8.6\\
%&6750,0.0,7&&9.3&9.1&8.4\\
%&6750,1.0,7&&9.1&9.1&8.4\\
%&6750,1.0,9&&8.9&9.1&8.4\\
%&6750,1.0,5&&9.3&9.1&8.4\\
%\hline
%\end{tabular}
%\end{minipage}
%\end{table}

The sensitivities of the permitted and forbidden lines to
the choice of model atmosphere are different. To illustrate
these sensitivities, we give in Table 3 the abundances derived from
R\,CrB's lines for a series of model atmospheres centred on 
Asplund et al.'s choice of $(T_{\rm eff}, \log g, \xi)$
= (6750, 0.5, 7). The carbon abundance from C\,{\sc i} lines is,
as expected from the insensitivity of the ratio of line to continuous
opacity to physical conditions, almost independent of the choice of
model. The abundance is also insensitive to the choice of the
microturbulent velocity $\xi$ for a weak line but not for a strong
line. In Table 3, the mean C abundance derived from weak and strong
permitted lines is given. The abundance from the forbidden
lines is insensitive to the choice of the surface gravity but
dependent on $T_{\rm eff}$. We note that the change $\pm$ 250 K in
$T_{\rm eff}$ leads to
a change in the carbon abundance of $\pm$ 0.2 dex. There is no
dependence on $\xi$. The $T_{\rm eff}$ and $\log g$ sensitivities
are not very different for the hottest stars (XX\,Cam and RY\,Sgr)
and the coolest (GU\,Sgr). 

To increase the carbon abundance from the 9850 \AA\ [C\,{\sc i}]
line to the input abundance ($\log\epsilon$(C) = 9.5), requires
that the $T_{\rm eff}$ of R\,CrB and SU\,Tau be raised about 
500 K, and 750 K, respectively. These increases are not
only outside the bounds considered acceptable by Asplund
et al., but they do not remove the carbon problem for the
permitted carbon lines. In addition,  they introduce
a  discrepancy between the
N abundance from the N\,{\sc i}  and CN lines. Also note that
the higher temperatures do not eliminate the abundance difference
of 0.7 dex from the 9850 \AA\ and 8727 \AA\ forbidden lines
of R\,CrB and SU\,Tau. In short, the [C\,{\sc i}] lines are
part of a now enlarged carbon problem.     

\begin{table*}
\centering
\begin{minipage}{55mm}
\caption{Carbon abundances for R\,CrB for various model atmospheres}
\begin{tabular}{cccccc} \hline
\multicolumn{1}{c}{}&\multicolumn{1}{c}{}&\multicolumn{1}{c}{}& \multicolumn{3}{c}{log$\epsilon$(C)} \\
\multicolumn{3}{c}{Model}& \multicolumn{1}{c}{C\,{\sc i}} &
\multicolumn{1}{c}{$[\rm C\,{\sc i}]$} &\multicolumn{1}{c}{$[\rm C\,{\sc i}]$}\\
\cline{1-3}
\multicolumn{3}{c}{$T_{\rm eff}$,log$g$,$\xi$}& \multicolumn{1}{c}{}&
\multicolumn{1}{c}{9850\AA}&\multicolumn{1}{c}{8727\AA}\\ \hline
&6750,0.5,7&&9.2&9.1&8.4\\
&6500,0.5,7&&9.2&8.9&8.2\\
&7000,0.5,7&&9.3&9.3&8.6\\
&6750,0.0,7&&9.3&9.1&8.4\\
&6750,1.0,7&&9.1&9.1&8.4\\
&6750,1.0,9&&8.9&9.1&8.4\\
&6750,1.0,5&&9.3&9.1&8.4\\
\hline
\end{tabular}
\end{minipage}
\end{table*}

%\begin{table}
%\centering
%\begin{minipage}{55mm}
%\caption{Carbon abundances for R CrB for various model atmospheres}
%\begin{tabular}{cccccc} \hline
%\multicolumn{1}{c}{}&\multicolumn{1}{c}{}&\multicolumn{1}{c}{}& \multicolumn{3}{c}{log$\epsilon$(C)} \\
%\multicolumn{3}{c}{Model}& \multicolumn{1}{c}{C\,{\sc i}} &
%\multicolumn{1}{c}{$[\rm C\,{\sc i}]$} &\multicolumn{1}{c}{$[\rm C\,{\sc i}]$}\\
%\cline{1-3}
%\multicolumn{3}{c}{$T_{\rm eff}$,log$g$,$\xi$}& \multicolumn{1}{c}{}&
%\multicolumn{1}{c}{9850\AA}&\multicolumn{1}{c}{8727\AA}\\ \hline
%&6750,0.5,7&&9.2&9.1&8.4\\
%&6500,0.5,7&&9.2&8.9&8.2\\
%&7000,0.5,7&&9.3&9.3&8.6\\
%&6750,0.0,7&&9.3&9.1&8.4\\
%&6750,1.0,7&&9.1&9.1&8.4\\
%&6750,1.0,9&&8.9&9.1&8.4\\
%&6750,1.0,5&&9.3&9.1&8.4\\
%\hline
%\end{tabular}
%\end{minipage}
%\end{table}

\section{Discussion}

Possible solutions to the carbon problem presented by the
C\,{\sc i} lines were 
discussed by Asplund et al. (2000 --see also Gustafsson \& Asplund 1996).
 Two proposed
solutions were suggested by Asplund et al.
as worthy of further consideration: the $gf$-values
of the C\,{\sc i} lines are in error, or theoretical model
atmospheres are a misrepresentation of the temperature
structure of real RCB stars. In addition to  commenting
on these solutions as they affect the forbidden lines,
we examine
 departures from LTE as they affect  
line formation. Finally, we discuss hand-crafted model atmospheres
including models
with a chromosphere (a temperature rise in the outer layers).

\subsection{The $gf$-values}

The $gf$-values of the [C\,{\sc i}] lines are known accurately from
theoretical calculations. 
%Differences at the 10 \% level may be
%debated but the large and different decreases in the $gf$-values
%of the 8727 \AA\ and 9850 \AA\ lines that are  needed
%to eliminate the carbon problem are outside the realm of
%possibility. 
Confirmation and extension of the carbon problem from the forbidden
lines strongly suggests that the $gf$-values of the permitted
lines can  not be primarily responsible for the problem.

\subsection{ Non-LTE Effects}

Non-LTE effects in neutral carbon atoms in RCB star atmospheres
were evaluated by Asplund \& Ryde (1996). Their  results
indicate that departures from LTE are confined to shallow
optical depths. Effects on the C\,{\sc i} lines are slight
because a  line and the continuum
 are formed deep in the atmosphere between
levels in the
carbon atom with small and very similar departures from LTE.
Gustafsson \& Asplund (1996) give the correction to LTE
abundance from C\,{\sc i} lines
 as less than 0.1 dex and suggest that $+$0.02 dex is a typical
value. Thus, non-LTE effects can not account for the
0.6 dex typical carbon problem arising from the permitted lines.
Departure
coefficients given by Asplund \& Ryde show that a non-LTE  correction
to the LTE abundance from a forbidden line  is also very small
and cannot be the source of the line's carbon problem.
These non-LTE calculations adopt a model computed in
LTE. 

Carbon problems may be reduced or even eliminated by invoking
a chromosphere.
A temperature rise at the top of the photosphere
is possible even in the absence of mechanical energy deposition.
This non-LTE effect  sometimes called the Cayrel mechanism
(Cayrel 1963) is discussed by Mihalas (1978).  To assess the
temperature rise for a RCB star due to the Cayrel mechanism,
non-LTE effects on the populations of the carbon atom were
included in the computation of the model atmosphere.
These non-LTE effects on the carbon level-populations were calculated 
using MULTI (Carlsson 1986), with the modelled carbon atom 
described in Asplund \& Ryde (1996). Realistic background fluxes, computed with 
the MARCS model atmosphere code, were included. Typical departure coefficients 
(b=n/n(LTE), where n is the occupation number density) of about b = 1 for the
lower C\,{\sc i} levels, b = 1.4 for intermediate ones and b = 0.6 for the upper ones 
were obtained. These figures were found to be characteristic of the depth 
interval $- 4 < \log \tau_{\mathrm{Ross}} < - 3$, with much smaller and opposite 
effects at larger optical depths.
These departures from LTE should mainly affect the opacities. C\,{\sc i} is also 
an important electron contributor, but our calculations show only an 
underionization of less than 10\% at shallow depths.

We then brought these results into the MARCS program
in order to include their effect  on  the
continuous opacities. This procedure is not fully self-consistent - the models 
are still LTE models - but the main effect of the departures on the opacities 
are modelled in a reasonable way. They essentially diminish the 
C\,{\sc i} opacities longwards of 6000 \AA, to 0.6 times the LTE value, 
and increase them shortwards of that wavelength to 1.4 times the LTE value. 
This was applied for the uppermost atmosphere with a decrease corresponding 
to our non-LTE results at greater depth.
A Cayrel effect resulted, as expected, with a heating of the upper layers, 
but only by about 1-6 K, the maximum occuring at log $\tau_{\mathrm{Ross}} = -2$. 

Thus, the resultant Cayrel effect is small for C\,{\sc i} in RCB stars with a 
temperature increase less than 10 K at the top of the photosphere. This 
micro-chromosphere does not alleviate the carbon problem! The small Cayrel 
effect is partly
linked to the fact that carbon atoms are 
 not totally dominating the scene. In fact at 
these relevant depths photoionization of carbon atoms
only contributes typically 50\% of the opacity. So, even drastic  
changes (at least reductions) to the carbon opacity
 only lead to moderate changes of fluxes 
and of the energy balance. Another circumstance is that the ultraviolet
flux in the 
surface layers is so weak that it does not seem to matter very much and is unlike 
the situation for hot stars and the H\,{\sc i} opacity, only a tiny fraction of 
the flux of typical RCB stars comes at energies higher than the ionization 
threshold, or even at energies able to photoionize from lower excited C\,{\sc i}
levels.

\subsection{Hand-crafted Models}

The fact that the carbon problems defined by the C\,{\sc i}
and the [C\,{\sc i}]
lines are very similar from star-to-star across the sample
of analysed RCB stars implies that the problems' resolution cannot depend
very sensitively
on a star's individual characteristics.
Departures of the atmospheric structure from that predicted
by a standard (MARCS) model must be very similar across the
sample of stars. If departures are attributable to deposition of
mechanical energy, the flux of mechanical energy cannot vary
widely across the sample. 

That the atmospheric structure is partly the culprit is suggested
by the observation that an RCB star in a decline shows a reduced
carbon problem. The case of V482\,Cyg was discussed in
Section 5.8. There, the carbon problem for the C\,{\sc i} lines
vanished. Note that the C\,{\sc i} lines are noticeably stronger in our
spectrum than in spectra taken at maximum light.
 The problem remains for the 9850 \AA\ forbidden line, but
this might be the result of overlying emission, as occurred in the
case of RY\,Sgr. (Unfortunately, the 8727 \AA\ line was not on the
recorded spectrum.) A similar strengthening of C\,{\sc i} lines
leading to elimination of the carbon problem but with retention
of the problem for the [C\,{\sc i}] 9850 \AA\ line
was seen by Rao et al. (1999) for R\,CrB in decline.

\subsubsection{The photospheric temperature gradient}

The carbon problem may be related to an incorrect representation by the
MARCS RCB star models of the
temperature gradient through the region of line formation; 
a steeper/shallower gradient produces a stronger/weaker
 line. In this connection, it
may be noted that  Lambert \& Rao's (1994) abundance analysis
used unblanketed model atmospheres (Sch\"{o}nberner 1975) and
found  fair agreement between the observed and calculated
equivalent widths of the C\,{\sc i} lines. Note that the small carbon problem of
0.2 dex  might be attributed to a combination of
identifiable uncertainties. The primary factor responsible for 
the negligible carbon problem
is that Sch\"{o}nberner's
unblanketed models have a shallower temperature gradient in the
line forming region than the
MARCS (line blanketed) RCB star models (Gustafsson \& Asplund 1996).

 Unblanketed models were
 calculated with the MARCS code. 
Our abundance analysis for R\,CrB using unblanketed MARCS models
shows that the carbon problems presented by
permitted and forbidden C\,{\sc i} lines are similar for
unblanketed and blanketed MARCS models. 
 The carbon abundance derived from [C\,{\sc i}] 9850 \AA\
line using unblanketed  models again agrees that from the 
 C\,{\sc i} lines. The large difference between the carbon abundance
derived from the [C\,{\sc i}] 9850 \AA\ line
 and the [C\,{\sc i}] 8727 \AA\ line using  
the  blanketed models is unaltered by using the unblanketed
models. The difference between the unblanketed MARCS and Sch\"{o}nberner's
models reflects differing  temperature gradients of the
models.

Asplund et al. describe hand-crafted adjustments to the line blanketed models that
reduce the photospheric temperature gradient and alleviate the carbon problem
presented by the C\,{\sc i} lines.
A shallower gradient (with respect to a blanketed MARCS model) 
 was qualitatively justified as arising from  
deposition of mechanical energy whose source may be the
sub-photospheric convection zone.
Examples of  temperature modified models, which alleviate the
carbon problem from C\,{\sc i} lines, were kindly provided by
Martin Asplund (private communication). Our analyses using these
models show that the carbon problem vanishes for the
9850 \AA\ [C\,{\sc i}] line as it does for the C\,{\sc i} lines, but
the 
8727 \AA\ [C\,{\sc i}] line continues to present a problem;
 the carbon abundance from 
8727 \AA\ [C\,{\sc i}] line is about 0.7 dex less than that from 
9850 \AA\ [C\,{\sc i}] line, a difference found also for 
unmodified line blanketed and unblanketed models.

\subsubsection{A chromosphere}

Addition of a temperature rise (`a chromosphere')
at the top of the photosphere may also modify the predicted
strengths of the  absorption lines in the emergent spectrum.
The chromosphere must be crafted 
by hand without theoretical guidance, but mindful of the discussion in
the previous section.

\begin{figure}
\epsfxsize=8truecm
\epsffile{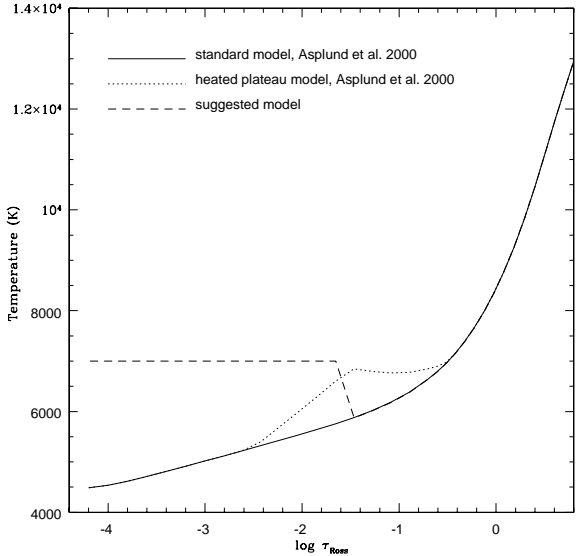}
\caption{Three $T$ $-$ $\tau_{\mathrm Ross}$ structures are shown - see key at the top
of the figure.  The solid line shows a MARCS model for  $T_{\rm eff}$ = 6750
and log$g$ = 0.5. The modified   $T$ $-$ $\tau$ structure with a heated plateau
devised by Asplund et al. (2000) to eliminate the carbon problem for
C\,{\sc i} lines is shown by the dotted line. Our model with a chromosphere
is shown by the dashed line.}
\end{figure}

Experiments were made by adding a chromosphere to a MARCS line-blanketed
 model.
Figure 5 shows one experiment appropriate for R\,CrB itself.
A temperature rise 
is introduced 
at $\log\tau_{\mathrm{Ross}} = -$1.6.
 A flat temperature profile of 7000 K in the interval
$\log\tau_{\mathrm{Ross}} \le -$1.6 and a LTE spectrum synthesis
% (with $n_e$ $\approx$ 2 $\times$ 10$^{11}$ cm$^{-3}$)
reproduce the profiles of the observed
 [C\,{\sc i}] 9850 \AA\ and [C\,{\sc i}] 8727 \AA\ absorption lines. 
The two lines now return the same C abundance which is equal to
the input abundance, or, the carbon problem vanishes for these lines. 
The  weaker
observed C\,{\sc i} lines are also reproduced with the input abundance.
Very strong C\,{\sc i} lines are predicted, as expected for an LTE
calculation, to have
emission cores.
The emission cores might be reduced
if the non-LTE effects were taken into consideration.
Quite obviously, the chosen chromosphere is not a unique solution. If the onset of
the sharp temperature rise is placed as in Figure 5, a chromospheric
temperature of 6750 to 7000 K  provides acceptable
solutions to the carbon problems. Similar chromospheres remove the
carbon problems for the other RCB stars. The chromospheric
temperature may scale with a star's effective temperature, but the range of
sampled effective temperatures is small. 

Addition of a chromosphere not only changes the predicted
strengths of the C\,{\sc i} and [C\,{\sc i}] lines  but also the
predicted strengths of other lines and, therefore, the derived
abundances. High-excitation lines such as the N\,{\sc i} line at 8729 \AA\
are so weakened by chromospheric emission that the  observed 
absorption line in XX\,Cam and R\,CrB cannot be fit whatever the
adopted nitrogen abundance. 
Although
careful crafting of the chromospheric temperature profile may
alleviate this difficulty, it seems likely that a non-LTE
synthesis may be necessary. One may need to adopt
spherical rather than a plane-parallel geometry. The chromosphere, if
it exists, is likely to be highly-structured and a far  cry from
the homogeneous plane-parallel (or spherical) layers assumed
for the models.

Although the artifice of a chromosphere serves to reduce or even
eliminate the carbon problems, its presence presents another
puzzle.
The energy flux needed to produce the extra heating
suggested here is substantial. Following Asplund et al. (2000),
we find it to be on the order of 10 -- 15 \% of the total
stellar flux. For comparison, the chromospheric--coronal heating of
normal late-type stars is limited to about 1 \% of the total
bolometric flux. A clue to the problem may lie in the very large
widths of RCB star lines -- compare the widths of lines (Figure 1 and 2)
in $\gamma$ Cyg and the RCB stars. These large widths imply a very turbulent
atmosphere and dissipation of energy. The possibility of an external 
radiation field generating the very large RCB star line widths needs to
be explored.

Emission lines appear when an RCB star goes into decline. The
emitting gas is not necessarily to be identified with the
hand-crafted chromosphere. Emission from a shell around the
star would dilute the photospheric absorption lines. 
Emission in the 9850 \AA\ and 8727 \AA\ lines might
weaken the absorption lines such that the emission is not
seen but the weakened absorption lines  return
different carbon abundances. The [C\,{\sc i}] lines
are seen in emission in spectra taken in decline (Rao \& Lambert
1993; Rao et al. 1999).
The measured fluxes in decline are possibly lower
limits to the fluxes at maximum light as the dust cloud
responsible for the decline presumably obscures parts of the
emitting shell.
 At maximum light, the cores of the strongest low
excitation absorption lines of singly-ionized metals appear
distorted by emission (Lambert, Rao \& Giridhar 1990; Rao et al. 1999).
It is just such lines which are prominent in the emission
line spectrum.  
 The emission line spectrum is of low
excitation and does not include the C\,{\sc i} lines.
Therefore, the emitting shell which may contribute to the
[C\,{\sc i}] problem is not a player in the C\,{\sc i}
problem.
In this context, it is relevant to recall the case of V482\,Cyg
observed below maximum light where
the carbon problem is absent for the
C\,{\sc i} lines but not for the [C\,{\sc i}] 9850 \AA\
line. Our speculation is that the emitting shell fills in the forbidden
line and the photospheric temperature gradient is steeper than 
at maximum light.

\section{Concluding remarks}

The intriguing carbon problem presented by the
RCB stars 
%and their C\,{\sc i} lines 
was discovered 
by Asplund et al. (2000) in an application of MARCS RCB star
models to analysis of C\,{\sc i} lines. Our contribution to the problem has been
to present and analyse observations of the [C\,{\sc i}] 8727 \AA\
and 9850 \AA\ lines, also with the MARCS models.
 The 9850 \AA\ line
presents a problem of similar magnitude (about 0.6 dex) 
%to that provided by the 
as C\,{\sc i} lines,
but the carbon problem presented
by the 8727 \AA\ line is more severe by about 0.7 dex than the original
problem (about 0.6 dex).
%discovered by Asplund et al. from
%their analyses of C\,{\sc i} lines. The original carbon problem has been
%given a new dimension.

The fact that the carbon problems defined by the C\,{\sc i}
and the [C\,{\sc i}]
lines are similar from star-to-star across the sample
of analysed RCB stars implies that the solution cannot depend
on a star's individual characteristics.
Departures of the atmospheric structure from that predicted
by a standard (MARCS) model must be  similar across the
sample of stars. If departures are attributable to deposition of
mechanical energy, the flux of deposited mechanical energy cannot vary
widely across the sample. 

Uncovering the carbon problem was a surprise. Confidence in model
atmospheres was shaken. Extension of  the problem to the [C\,{\sc i}]
lines is also discomforting.
In spite of the fact that the RCB stars in several ways are very
complex, they have the one property that should make the analysis
of C\,{\sc i} lines in principle simpler than for almost all other
stars. This as the fact that the lines are due to the same element is the dominant
opacity source and lines and opacity originate from levels of 
similar character and excitation. This provides a test, that is not offered for
other stars. This test, however, fails.

 Perhaps, the problems are restricted
to very peculiar stars such as the rare RCB stars. But this
supposition deserves to be tested. Luminous normal (i.e., H-rich) supergiants 
of the temperature of the RCB stars are the stars from
which chemical compositions of external galaxies are now being
derived. Are their atmospheric structures  reliably simulated
by our codes? Or is there a problem analagous to the carbon problem
awaiting discovery?

This research has been supported in part by The Robert A. Welch
Foundation through a grant to DLL.


\begin{thebibliography}{99}

\bibitem{} Allende Prieto, C., Lambert, D. L., Asplund, M., 2002, ApJ, 573, L137
\bibitem{} Asplund, M., Gustafsson, B., Kiselman, D., Eriksson, K., 1997, A\&A, 318, 521
\bibitem{} Asplund, M., Gustafsson, B., Lambert, D. L., Rao, N. K., 2000, A\&A, 353, 287
\bibitem{} Asplund, M., Ryde, N., 1996, in:
     Hydrogen deficient stars, Jeffery C.S., Heber U.
    (eds.). ASP conf. series vol. 96, p. 57
\bibitem{} Bauschlicher, C. W., Jr., Langhoff, S. R., Taylor, P. R., 1988, ApJ, 332, 531
\bibitem{} Bakker, E. J., Lambert, D. L., 1998, ApJ, 502, 417
\bibitem{} Carlsson, M., 1986, Uppsala Astronomical Report, No. 33
\bibitem{} Cayrel, R., 1963,  C. R. Acad. Sci., 257, 3309
\bibitem{} Cottrell, P. L., Lambert, D. L., 1982, ApJ, 261, 595
\bibitem{} Davis Sumner, P., Phillips John, G., 1963, The red system (A$^2$$\Pi$$-$X$^2$$\Sigma$) of the CN molecule,
     Berkeley Analyses of Molecular Spectra, Berkeley: University of California Press.
\bibitem{} Galavis, M. E., Mendoza, C., Zeippen, C. J., 1997, A\&AS, 123, 159
\bibitem{} Gustafsson, B., Asplund, M., 1996, in:
     Hydrogen deficient stars, Jeffery C.S., Heber U.
    (eds.). ASP conf. series vol. 96, p. 27
\bibitem{} Gustafsson, B., Bell, R. A., Eriksson, K., Nordlund, A., 1975, A\&A, 42, 407
\bibitem{} Hibbert, A., Bi\'{e}emont, E., Godefroid, M., Vaeck, N., 1993, A\&AS, 99, 179
\bibitem{} J{\o}rgensen, U. G., Larsson, M, 1990, A\&A, 238, 424
\bibitem{} Lambert, D. L., Rao, N. K., Giridhar, S., 1990, JAA, 11, 475
\bibitem{} Lambert, D. L., Rao, N. K., 1994, JAA, 15, 47
\bibitem{} Lawson, W. A., Cottrell, P. L., Clark, M., 1991, MNRAS, 251, 687
\bibitem{} Liu, X. W., Barlow, M. J., Danziger, I. J., Clegg, R. E. S., 1995, MNRAS, 273. 47
\bibitem{} Luck, R. E., Lambert, D. L., 1981, 245, 1018
\bibitem{} Luo, D., Pradhan, A. K., 1989, JPhB, 22, 3377
\bibitem{} Mihalas, D. 1978, Stellar Atmospheres, Freeman \& Co.:San Francisco
\bibitem{} Moore, Ch. E.: 1972, A Multiplet Table of Astrophysical Interest,
    NSRDS - NBS 40, Washington
\bibitem{} Moore, Ch. E.: 1993, Tables of Spectra of Hydrogen, Carbon, Nitrogen, and
    Oxygen Atoms and Ions, Editor: Jean W. Gallagher, CRC Series in Evaluated Data
    in Atomic Physics, CRC press
\bibitem{} Nave, G., Johansson, S., Learner, R. C. M., Thorne, A. P., Brault, J. W., 1994, ApJS, 94, 221
\bibitem{} Rao, N.K., Lambert, D.L., 1993, AJ, 105, 1915
\bibitem{} Rao N. K., Lambert D. L.,1996, in:
    Hydrogen deficient stars, Jeffery C.S., Heber U.
    (eds.). ASP conf. series vol. 96, p. 43
\bibitem{} Rao N. K., Lambert D. L., 1997, MNRAS, 284, 489
\bibitem{} Rao N. K., Lambert D. L., Adams, M. T., Doss, D. R., Gonzalez, G., Hatzes, A. P.,
    James, R., Johns-Krull, C. M., Luck, R. E., Pandey, G., Reinsch, K.,
    Tomkin, J., 1999, MNRAS, 310, 717
\bibitem{} Sch\"{o}nberner, D., 1975, A\&A, 44, 383
\bibitem{} Searle, L., 1961, ApJ, 133, 531
\bibitem{} Seaton, M. J., Yan, Y., Mihalas, D., Pradhan, A. K., 1994, MNRAS, 266, 805
\bibitem{} Swensson, J. W., Benedict, W. S., Delbouille, L., Roland, G., 1970, The
    Solar Spectrum from $\lambda$7498 to $\lambda$12016. A Table of Measures and
    Identifications. Mem. Soc. R. Sci. Li\`{e}ge, Spec. Vol. No. 5
\bibitem{} Tull, R. G., MacQueen, P. J., Sneden, C. and Lambert, D. L., 1995, PASP,
    107, 251
\bibitem{} Wiese, W. L., Fuhr, J. R., Deters, T. M., 1996, Journal of Physical and Chemical Reference Data, Monograph No. 7

\end{thebibliography}
\end{document}